\documentclass[aps,pra,amsmath,amssymb,superscriptaddress,showpacs,reprint]{revtex4-1}
\usepackage{graphicx}
\usepackage{amsmath}
\usepackage{epsfig}
\usepackage{bm}
\usepackage{dcolumn}
\usepackage{amsbsy}
\usepackage[colorlinks=true,linkcolor=blue]{hyperref}
\begin{document}

\preprint{AIP/123-QED}

\title{Temporal steering in four dimensions with applications to coupled qubits and magnetoreception.}


\author{Huan-Yu Ku}
\affiliation{Department of Physics National
Cheng Kung University, Tainan 701, Taiwan}
\author{Shin-Liang Chen}
\affiliation{Department of Physics National
Cheng Kung University, Tainan 701, Taiwan}
\affiliation{Naturwissenschaftlich-Technische Fakult\"at,
Universit\"at Siegen, Walter-Flex-Str. 3, D-57068 Siegen, Germany}
\author{Hong-Bin Chen}
\affiliation{Department of Physics National
Cheng Kung University, Tainan 701, Taiwan}
\author{Neill Lambert}
\email{nwlambert@riken.jp}
\affiliation{RIKEN, 351-0198 Wako-shi, Japan
}
\author{Yueh-Nan Chen}
\email{yuehnan@mail.ncku.edu.tw}
\affiliation{Department of Physics National
Cheng Kung University, Tainan 701, Taiwan}
\affiliation{RIKEN, 351-0198 Wako-shi, Japan
}
\affiliation{National Center for Theoretical Sciences, Hsinchu 300, Taiwan
}
\author{Franco Nori}
\affiliation{CEMS,RIKEN, 351-0198 Wako-shi, Japan
}
\affiliation{Department of Physics, The University of Michigan, Ann Arbor, Michigan 48109-1040, USA}
\date{\today }

\begin{abstract}
Einstein-Podolsky-Rosen (EPR) steering allows Alice to remotely prepare a state in some specific bases for Bob through her choice of measurements. The temporal analog of EPR steering, temporal steering, also reveals the steerability of a single system between different times. Focusing on a four-dimensional system, here we investigate the dynamics of the temporal steering measures, the temporal steering robustness, using five mutually unbiased bases. As an example of an application, we use these measures to examine the temporal correlations in a radical pair model of magneto-reception. We find that, due to interactions with a static nuclear spin, the radical pair model exhibits strong non-Markovianity.
 
\end{abstract}

\maketitle

\affiliation{Department of Physics and National Center for Theoretical Sciences, National
Cheng Kung University, Tainan 701, Taiwan}

\affiliation{Department of Physics and National Center for Theoretical Sciences, National
Cheng Kung University, Tainan 701, Taiwan}

\affiliation{Department of Physics and National Center for Theoretical Sciences, National
Cheng Kung University, Tainan 701, Taiwan}

\section{introduction}
Quantum steering~\cite{Schrodinger1,Reid1,Wiseman1,Cavalcanti01} is an intriguing phenomenon wherein one party can \emph{remotely steer} the quantum state of another party through their choice of measurements.
Remarkably, there exists a hierarchy relation between steering, Bell nonlocality, and entanglement. That is, states which are Bell nonlocal are also steerable, and all steerable states are entangled, but not vice versa~\cite{Wiseman1,Quintino15}. Numerous applications of steering have been considered, such as the connection to one-side device independent quantum key distribution~\cite{1SDQKD,PRX}, a geometrical representation of steering~\cite{Jevtic14}, the correspondence with measurement incompatibility~\cite{Uola14,Quintino14,Uola15}, steering beyond quantum theory~\cite{Sainz15}, multipartite steering~\cite{Armstrong15,Li_Che_Ming,Cavalcanti15}, etc. In addition, there have been many efforts at quantifying steering~\cite{Paul_Skrzypczyk1,Piani,Kogias15,PRX,Costa2016,Cavalcanti16}. In addition, many experiments exhibiting the reality of steering have also been performed~\cite{Wittmann,Smith_Devin_H,Saunders,Li_Che_Ming}.

A range of different types of quantum correlations also appear when measuring a single system at different times. For example, the Leggett-Garg (LG) inequality~\cite{Leggett,LG_report}, a temporal analog of Bell's inequality, based on the assumption of macroscopic realism, relies on combining two-time correlation functions~\cite{Emary2012,Lambert2010}. 
Similarly, other types of temporal correlations have been proposed and investigated, including quantum entanglement in time, temporal nonlocality, and bounding temporal quantum correlations~\cite{Brukner04,Fritz10,Costantino13,Costantino14}.
Motivated by the correspondence between Bell's nonlocality and the LG inequality, a temporal analog of steering was proposed by Chen \emph{et al.}~\cite{Yueh-Nan1,Chen-S-L1,Che-Ming1}. Focusing on a single system transmitted from Alice to Bob, temporal steering demonstrates Alice's influence on Bob via her choice of measurements. Temporal steering is related to quantum key distribution~\cite{Yueh-Nan1,Chen-S-L1,Bartkiewicz15,Che-Ming1}, measurement incompatibility~\cite{Karthik15}, and quantum non-Markovianity~\cite{Chen-S-L1}. The first experiment showing temporal steering has also recently been reported by Bartkiewicz \emph{et al.}~\cite{Bartkiewicz16}.

Although some works concerning temporal steering have been proposed, research on  temporal steering in higher dimensions is still lacking.
Here, we introduce a new quantifier, temporal steering robustness, in analogy with  spatial steering robustness~\cite{Piani}.
Then, we move on to considering the temporal steering robustness of four dimensional systems. As examples, we first consider two coupled qubits, and construct its temporal assemblage using five mutually unbiased bases (MUBs)~\cite{Klappenecker03}. Second, we consider the radical-pair model, a ``toy model" used to describe the sensitivity of certain chemical reactions to magnetic fields, and which is one of the candidate models for the origin of avian magnetoreception.
Finally, we investigate the non-Markovianity of the dynamics of electrons in the radical pair model, as revealed by nonmonotonic temporal steering.

\section{Temporal steering and how to quantify it}
\subsection{Formulation of temporal steering}
First, let us briefly review the concept of temporal steering. Alice performs a measurement, which can be described by a set of positive-operator valued measures (POVMs) $\{E_{a|x}\}$, with measurement choice $x$ on an initial state $\rho_0$ at time $t=0$. After the measurement, she obtains an outcome $a$ and a postmeasurement state $\hat{\sigma}_{a|x}(t=0)=M_{a|x}\rho_{0}M^{\dagger}_{a|x}/p(a|x)$, where $p(a|x)=\text{tr}(M_{a|x}\rho_{0}M^{\dagger}_{a|x})$, with $M^{\dagger}_{a|x}M_{a|x}=E_{a|x}$. After that Alice sends the state $\hat{\sigma}_{a|x}(t=0)$ to Bob through a quantum channel $\Lambda$, in which a unitary evolution or environment-induced noise may take place. After the transmission, Bob receives the assemblage $\hat{\sigma}_{a|x}(t) = \Lambda[\hat{\sigma}_{a|x}(t=0)]$ at time $t$.

To verify whether Alice's choice of measurement influences Bob's received state, Bob checks whether the assemblage ${\sigma}_{a|x}(t) := p(a|x)\sigma_{a|x}(t)$ can be written in a hidden-state form:
\begin{equation}
\label{temporal local hidden varable}
\sigma_{a|x}(t)~=~\sigma_{a|x}^{\textrm{T,US}}~=~
\sum_{\lambda}P(\lambda)~P(a|x,\lambda)~\sigma_{\lambda}.
\end{equation}
If it is the case, Bob would think that the probability distribution $P(\lambda)$ can be reconstructed from Alice's measurement setting $x$ and the outcome $a$. In addition, he would also think that the states he receives are predetermined by $\sigma_\lambda$ during each round of the experiment, and not actually influenced by Alice's measurement choice. Thus all Alice has to do is use her knowledge of the probability distribution $\lambda$ and $P(a|x,\lambda)$ to construct her measurement results. What Bob receives is the statistical average of the state of Eq.~(\ref{temporal local hidden varable}). Conversely, if it is not the case that his assemblage can be written in a hidden-state form, he convinces himself that the state he receives is actually influenced by Alice's choice of measurement.


In Ref.~\cite{Piani}, Piani and Watrous introduced a quantifier of steering --- steering robustness, the minimum noise needed to destroy the steerability of the assemblage. Here, we show that there also exists a temporal analog of steering robustness --- temporal steering robustness (TSR), that can serve as a quantifier of temporal steering.

Similar to the steering robustness, the temporal steering robustness is defined as the minimum noise needed to destroy the temporal steerability of the temporal assemblage:
\begin{equation}
\begin{aligned}
\text{TSR}:=\text{min} ~~&t\geq0 \\
\text{subject to }~~ &\left\{\frac{\sigma_{a|x}^\text{T}+t~\tau_{a|x}}{1+t}\right\}_{a,x}
~~\text{temporal unsteerable},\\
&\{\tau_{a|x}\}_{a,x}:\text{an assemblage}.
\end{aligned}  \label{TSR}
\end{equation}
Following the procedure in Ref.~\cite{Piani}, the condition~(\ref{TSR}) can also be written as an semidefinite programming (SDP) optimization problem:
\begin{equation}
\begin{aligned}
\text{TSR}=\text{min} ~~&\text{tr}\sum_{\lambda}\sigma_{\lambda}-1 \\
\text{subject to }
~~&\sum_{\lambda}D_{\lambda}(a|x)\sigma_{\lambda}\geq\sigma_{a|x}
~~&&\forall~a,x\\
&\sigma_{\lambda}\geq 0 &&\forall ~\lambda,
\end{aligned}  \label{TSR_SDP}
\end{equation}
where $\sigma_{\lambda}=(1+t)\sigma^{\textrm{T},\text{US}}_{a|x}$ and $D_{\lambda}(a|x)=\delta_{a,\lambda(x)}$~\cite{Chen-S-L1,Paul_Skrzypczyk1} is the deterministic value of the single-party conditional probability distributions $P(a|x,\lambda)$.
In the following section, we will use the temporal steering robustness to realize the temporal correlation in higher-order system for some specific quantum channel.


\section{Temporal Steering in Systems with Dimension  \texorpdfstring{$\bm{d=4}$}{d=4}}
\subsection{Two qubits coherently coupled with each other}
In this section, we examine the dynamics of the temporal steerability of a system composed of two qubits coherently coupled with each other, given by the interaction Hamiltonian $H=g(\sigma_1^+\sigma_2^- + \sigma_1^-\sigma_2^+)$, where $g$ is the coupling strength between the two qubits, and $\sigma_i^+$ and $\sigma_i^-$ are the raising and lowering operators of the $i$th qubit, respectively. In addition, each qubit is subject to a Markovian decay process. The evolution of the entire system is expressed by the master equation with Lindblad form~\cite{lindblad1976}
\begin{equation}
\dot{\rho}=\frac{1}{i\hbar}[H,\rho]+\sum_{i=1}^{2}\frac{\gamma}{2}(2\sigma_{i}^{-}\rho\sigma_{i}^{+}-\sigma_{i}^{+}\sigma_{i}^{-}\rho-\rho\sigma_{i}^{+}\sigma_{i}^{-}),
\end{equation}
where $\gamma$ is the decay rate.
Mathematically, we can treat the two qubits as a single four-dimensional system, i.e., $|gg\rangle\equiv|1\rangle,~~|ge\rangle\equiv|2\rangle,~~|eg\rangle\equiv|3\rangle$, and $|ee\rangle\equiv|4\rangle$, for which the maximum number of MUBs measurement is five. The set of $5$ MUBs is denoted by {$M_{a|x}=|\phi_{a|x}\rangle\langle\phi_{a|x}|$}~\cite{Wiesniak}, as detailed in Appendix A.

\begin{figure}
\includegraphics[width=8cm]{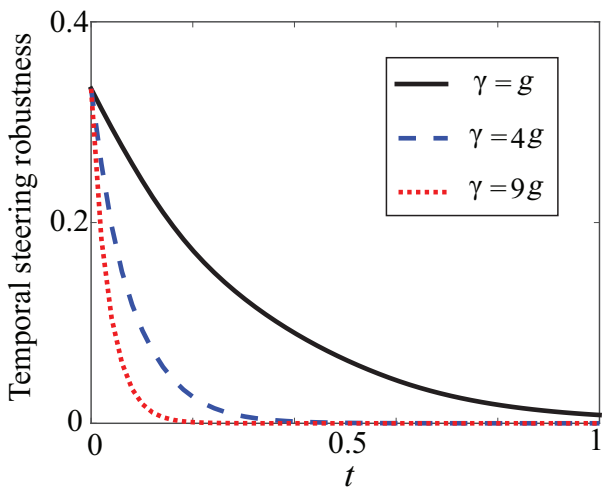}
\caption{(Color online) The dynamics of temporal steering robustness (two measurement settings) for two coherently coupled qubits with different decay rates $\gamma$.
The black-solid, blue-dashed, and red-dotted curves represent $\gamma=g, 4g,$ and $9g$, respectively. Here, $t$ is in units of $\gamma^{-1}$.
The initial state is in the maximally-mixed state $\rho(0)=\openone/4$.
}
\label{TSR_mixed_gamma_1_4_9_setting_12_12}
\end{figure}

\begin{figure}
\includegraphics[width=8cm]{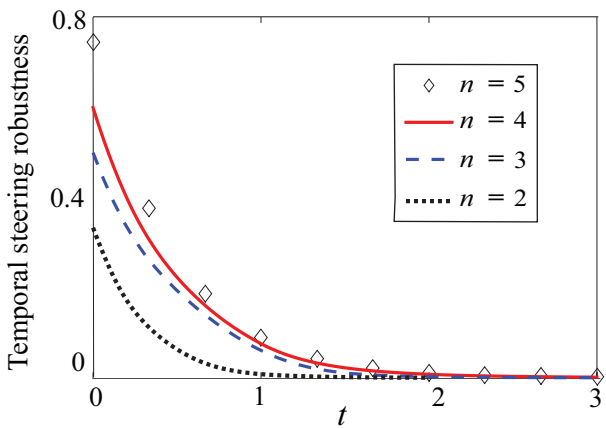}
\caption{The dynamics of temporal steering robustness of two qubits with different numbers of measurement settings $n$. The decay rate is set as $\gamma=g$, and the initial state is in the maximally-mixed state $\rho(0)=\openone/4$. Here, we compare the results of four kinds of measurement settings ($n=2$ to $5$). For example, when $n=3$, $\{M_{a|x}\} = \{M_{a|1}, M_{a|2}, M_{a|3}\}$. We see that the temporal steering robustness increases with the number of measurement settings, due to the intrinsic definition of the measure of (temporal) steerability~\cite{Paul_Skrzypczyk1,Piani,Chen-S-L1}.
}
\label{k_5_g_1_gamma_1}
\end{figure}

We assume that the initial state of the two-qubit system is the maximally-mixed state $\rho(0)=\openone/4$, where $\openone$ is the identity matrix. The postmeasurement state $\sigma_{a|x}(t)=M_{a|x}\rho(0)M_{a|x}/p(a|x)$ can be obtained straightforwardly. Figure~\ref{TSR_mixed_gamma_1_4_9_setting_12_12} shows the dynamics of the temporal steering robustness with two measurement settings ($n=2$ and choosing the measurement settings $x=1,2$) with different decay rates $\gamma$. In Fig.~\ref{k_5_g_1_gamma_1}, we compare the dynamics of temporal steering robustness for different numbers of measurement settings ($n=2$ to $n=5$, and choosing the measurement setting $x=1,...,n$ for each curve). We can see that the temporal steering robustness increases when the number of measurement settings increases, as expected from the original definition of the temporal steering robustness in Eq.~(\ref{TSR_SDP}).

\subsection{Temporal Steering Robustness of the Radical Pair}

\begin{figure}
\includegraphics[width=8cm]{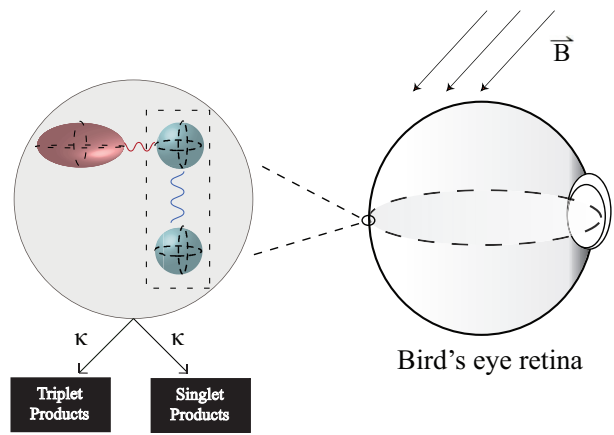}
\caption{(Color online) Schematic diagram of the radical-pair model. The radical-pair mechanism for avian navigation can explain some of the features of behavioral experiments of European robins~\cite{Erik,Andrea,RitzS1}. It is thought that it may occur within certain cryptochrome proteins residing in the eye. The simplest radical-pair toy-model is composed of two electrons and a nucleus, coupled to one of the electrons with the hyperfine interaction.
The singlet and triplet states of the two electrons in the radical pair inter-convert due to a combination of the Zeeman splitting due to the geomagnetic field, and an anisotropic nuclear hyperfine interaction.
At later times, the singlet and triplet states decay into chemical products, dependent on their spin nature, which we track with the ancilla shelving states $S$ and $T$, respectively.
}
\label{rp}
\end{figure}

The mechanism by which birds and other animals navigate using the geomagnetic field is still unclear. Among various proposals, the radical-pair model has received considerable attention due to its ability to predict many of the behavioral features seen in experiments and its uniquely quantum features~\cite{Erik}. In addition, radical-pair reactions are known to occur within the biological photoreceptor cryptochrome~\cite{Mouritsen1,Andrea}, perhaps leading to a biologically detectable signal.
In the traditional ``toy-model" of this process, a radical pair within or attached to the cryptochrome is formed when an electron is excited from a donor to a receptor molecule, which thus hosts spatially-separated electrons in a spin-singlet or triplet state.
The electron pair then evolves coherently between these states, under the influence of the geomagnetic field and the hyperfine interactions with the host nuclei~\cite{RitzS1,Neill}. At a later time, the singlet-triplet conversion leads to different chemical reaction products that could lead to a biologically-detectable signal. Figure~\ref{rp} depicts the basic concept of the radical-pair model.  Of course, in reality the chemical-process may be much more complicated than this toy-model suggests, but it is helpful to consider such a model because of its simplicity and intuitive ability to explain some behaviorial features. Despite this simplicity, here we find that the analysis of higher-dimensional steering in this model reveals some surprising and counterintuitive features.

The simplest radical-pair model contains two electrons and one nuclear spin~\cite{Erik}. The nucleus interacts with only one of the electrons, while the other is free. The hyperfine interaction between the nucleus and the electron together with the Zeeman effect induce the interconversion between the singlet and triplet states. For the radical-pair model to be sensitive to the angle of the external geomagnetic field, the hyperfine coupling tensor must be anisotropic.
The anisotropic hyperfine tensor between the nuclear spin and electron-1 can be written as $\mathbf{A}=\rm{diag}(\mathrm{A}_{\mathrm{x}},\mathrm{A}_{\mathrm{y}},\mathrm{A}_{\mathrm{z}})$. Here, we consider two kinds of anisotropic hyperfine tensors $\mathrm{A}_\mathrm{x}=\mathrm{A}_\mathrm{y}=0$, $|\mathrm{A}_\mathrm{z}|=10^{5}$ and $\mathrm{A}_\mathrm{x}=\mathrm{A}_\mathrm{y}=\mathrm{A}_\mathrm{z}/2$ with $|\mathrm{A}_\mathrm{z}|=10^{5}$ m eV ~\cite{Erik,Neill2,Pauls}. The Zeeman effect is included due to the coupling between the magnetic field and the electrons. 
The Hamiltonian of the entire system is
\begin{equation}
H=\sum_{i=1}^{2}\gamma \mathbf{B} \cdot  \mathbf{S}_{i}+\mathbf{I}\cdot \mathbf{A}  \cdot \mathbf{S}_{1},
\end{equation}
where $\mathbf{S}_{i}\equiv(\sigma_{\mathrm{x},i},\sigma_{\mathrm{y},i},\sigma_{\mathrm{z},i})$ are the electron spin operators ($i=\mathrm{1}$,2) with Pauli matrices $\sigma$, $\mathbf{I}$ is the spin operator for the nucleus, and $\mathbf{B}$ is the magnetic field. Here, $\gamma=\frac{1}{2}\mu_{B}g_{s}$ is the gyromagnetic ratio with $\mu_{B}$ being the Bohr's magneton and $g_{s}=2$ being the magnetic moment~\cite{Neill2}. The magnetic field for the two electrons and the nucleus can be generally described by
\begin{equation}
\mathbf{B} =B_{0}(\cos \phi \sin \theta,\sin \phi \sin \theta , \cos \theta),
\end{equation}
where $B_{0}$ = 47 $\mu$T is the intensity of the Earth's magnetic field. Without loss of generality, an axial symmetry is usually assumed: $\phi=0$ and $\theta \in [0,\pi/2]$.

To mimic the process that the singlet and triplet states decay to the chemical compounds, we additionally add two ancilla-shelving systems (called $S$ and $T$) to the Hilbert space to keep track of the population decay into singlet and triplet products, respectively. These are not physical systems but just mathematically convenient to aid in tracking the change in population. One can also adopt other approaches, which are typically more numerically conservative, but here it is convenient as we wish to investigate the temporal dynamics of the electron-spin systems without loss of population, which we can do by tracing out the ancillas. This corresponds to postselecting on populations which have not decayed.  Of course, if one cares about the magnitude of a signal corresponding to the decay processes, one should investigate these populations directly.

Later, we will use a master equation with the Lindblad terms to describe the Markovian decay process from the singlet state, as recorded by the ancilla $S$, as well as from the triplet state, as recorded by the ancilla $T$. The bases of every element of our system are as follows: First, the bases of the electron pair are defined as $\{|s\rangle,|t_{0}\rangle,|t_{-1}\rangle,|t_{+1}\rangle\}$, with $|s\rangle$ and $\{|t_{i}\rangle\}_{i=-1,0,1}$ being that singlet and triplet states, respectively. Second, $|\uparrow\rangle$ and $|\downarrow\rangle$ are the bases describing the nuclear-spin states. Finally, $\{|S_{j}\rangle\}$ and $\{|T_{j}\rangle\}$ (where $j=0,1$) are states of the ancilla $S$ and ancilla $T$, respectively, with $j=0$ describing the subspace where the system has not decayed, and $j=1$ the subspace where it has.
With the above definitions, we can now define the projection operators as $P_{s,\uparrow}=|s,\uparrow,S_{1},T_{0} \rangle \langle s,\uparrow,S_{0},T_{0}|$, $P_{t_{0},\uparrow}=|t_0,\uparrow,S_{0},T_{1} \rangle \langle t_{0},\uparrow,S_{0},T_{0}|$, $P_{t_{-1},\uparrow}=|t_{-1},\uparrow,S_{0},T_{1} \rangle \langle t_{-1},\uparrow,S_{0},T_{0}|$, and $P_{t_{+1},\uparrow}=|t_{+1},\uparrow,S_{0},T_{1} \rangle \langle t_{+1},\uparrow,S_{0},T_{0}|$. 
The projective operators describe the spin-selective recombination into the chemical compounds (ancilla $S$ and ancilla $T$ states). We also consider additional environmental noise described by the standard Lindblad formalism~\cite{Erik,Bandyopadhyay1}. The dynamics of the density matrix is obtained by solving the following master equation
\begin{equation}
\begin{aligned}
\dot{\rho}=&\frac{1}{i\hbar}[H,\rho]+\kappa\sum_{i}^{8}\bigg[P_{i}\rho P_{i}^{\dagger}-\frac{1}{2}(P_{i}^{\dagger}P_{i}\rho+\rho P_{i}^{\dagger}P_{i})\bigg]+\\
&\Gamma \sum_{i=1}^{2}\bigg[\sigma_{\mathrm{z},i}\rho\sigma_{\mathrm{z},i}^{\dagger}-\frac{1}{2}(\sigma_{\mathrm{z},i}^{\dagger}\sigma_{\mathrm{z},i}\rho+\rho\sigma_{\mathrm{z},i}^{\dagger}\sigma_{\mathrm{z},i})\bigg],
\end{aligned}
\end{equation}
where $\sigma_{z,i}$ are the Lindblad operators of the two electrons. Here, we assume that all the singlet and triplet recombination operators have the same decay rate $\kappa=10^{4}$ s$^{-1}$, and $\Gamma=10^{3}$ s$^{-1}$ is the rate of decoherence of each electron. The value $\kappa=10^{4}$ s$^{-1}$ is chosen as it is the one thought to explain certain experimental results in which a small oscillating magnetic field can disrupt the European Robins' ability to navigate~\cite{RitzS2,RitzS3,Erik}. An implication of these results is that the decoherence time of the radial-pair model could of the order of 100 $\mu$s or more~\cite{Erik}.

\begin{figure}
\includegraphics[width=8.7cm]{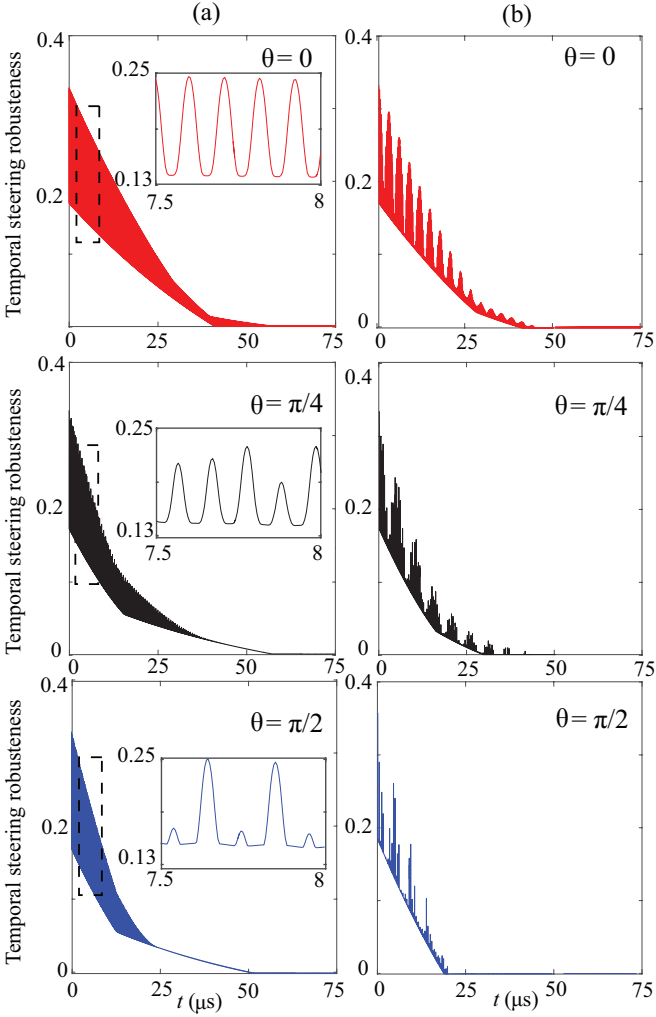}
\caption{(Color online) The dynamics of temporal steering robustness (two measurement settings, $M_{a|1}$ and $M_{a|2}$) of the radical-pair model.
The red, black and blue solid curves represent the results of the angle $\theta=0$, $\theta=\pi/4$, and $\theta=\pi/2$, between the magnetic field and the radical-pair, respectively.
In column~(a), we set the anisotropic tensor: $\mathrm{A}_\mathrm{x}=\mathrm{A}_\mathrm{y}=0$, $\mathrm{A}_\mathrm{z}=|10^{5}|$ meV~\cite{Erik,Neill2,Pauls}. The times when the signals vanish, for the red, black, and blue solid curves curves are $56$ $\mu\textbf{s}$, $53$ $\mu\textbf{s}$, and $50$ $\mu\textbf{s}$ , respectively. In column~(b), we set the anisotropic tensor: $\mathrm{A}_\mathrm{x}=\mathrm{A}_\mathrm{y}=\mathrm{A}_\mathrm{z}/2$ with $|\mathrm{A}_\textrm{z}|=10^{5}$ meV. The times when the signals vanish, for the red, black and blue solid curves are $45$ $\mu\textbf{s}$, $41$ $\mu\textbf{s}$, and $20$ $\mu\textbf{s}$, respectively. The dynamics of the temporal steering robustness obviously depends on the angle $\theta$ between the magnetic field and the radical-pair in this simplest model.
}
\label{TSR_mixed_setting_12}
\end{figure}

\begin{figure}
\includegraphics[width=8.7cm]{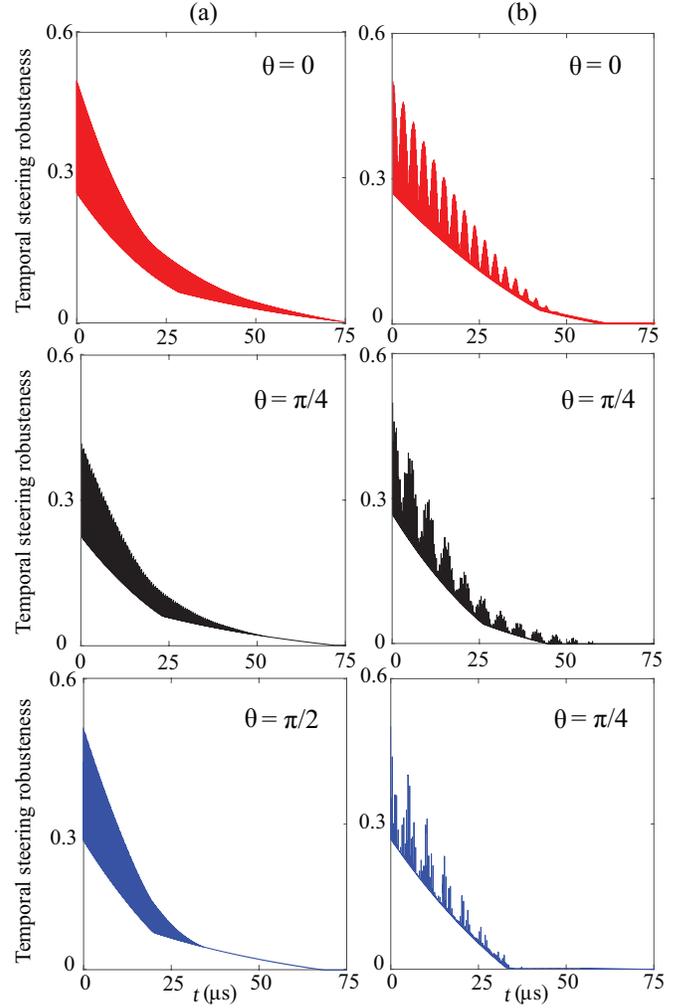}
\caption{(Color online) The dynamics of temporal steering robustness (three measurement settings, $M_{a|1}$, $M_{a|2}$ and $M_{a|3}$) of the radical-pair model. The red, black, and blue solid curves show the results for the angles $\theta=0$, $\theta=\pi/4$, and $\theta=\pi/2$, between the magnetic field and the radical-pair, respectively. The difference between Fig.~\ref{TSR_mixed_setting_12} and~\ref{TSR_mixed_setting_123} is the number of the measurement settings, $n_{x}$. In column~(a), the times when the signals vanish, for the red, black, and blue solid curves are  $75$ $\mu\textbf{s}$, $71$ $\mu\textbf{s}$, and $69$ $\mu\textbf{s}$, respectively. In column (b), the times when the signals vanish, for the red, black, and blue solid curves are $62$ $\mu\textbf{s}$, $57$ $\mu\textbf{s}$, and $34$ $\mu\textbf{s}$, respectively.
}
\label{TSR_mixed_setting_123}
\end{figure}

Previous works~\cite{Erik,Pauls,Cai} have looked at the behavior of the entanglement between the free electron and the electron coupled with the nucleus. Here, we are primarily interested in the temporal quantum correlations of the two-electron system at different times. Also, we assume that the initial state of the entire system (the two electrons, the nuclear spin, and the ancillas $S$ and $T$) is $\rho(t=0^{-})=\frac{1}{8}\times \openone\otimes |S_{0}\rangle \langle S_{0}|\otimes |T_{0}\rangle\langle T_{0}|$, where $\frac{1}{8}\times\openone$ is the maximally-mixed state of the two electrons and nuclear spin~\cite{Cai}. The five MUBs measurements are performed on the two-qubit system at time $t=0$, producing the temporal state assemblage $\sigma_{a|x}(t)$, and the dynamics of the temporal steering robustness can then be obtained. In Figs.~\ref{TSR_mixed_setting_12} and \ref{TSR_mixed_setting_123}, we plot the dynamics of the temporal steering robustness with two ($n=2$ and choosing the measurement setting $x=1$, 2) and three ($n=3$ and choosing the measurement setting $x=1$, 2, 3) measurement settings, respectively. Here, we can see that the dynamics of the temporal steering robustness is clearly dependent on the orientation $\theta$. While it is hard to state a strong connection between such temporal quantum correlations and the functionality of the avian compass, in the next section we will argue that these results imply a counterintuitive appearance of non-Markovianity in this model, easy to miss without looking at a quantity like the temporal steering robustness.

\subsection{The non-Markovianity of the Radical Pair}

\begin{figure}
\includegraphics[width=8.8cm]{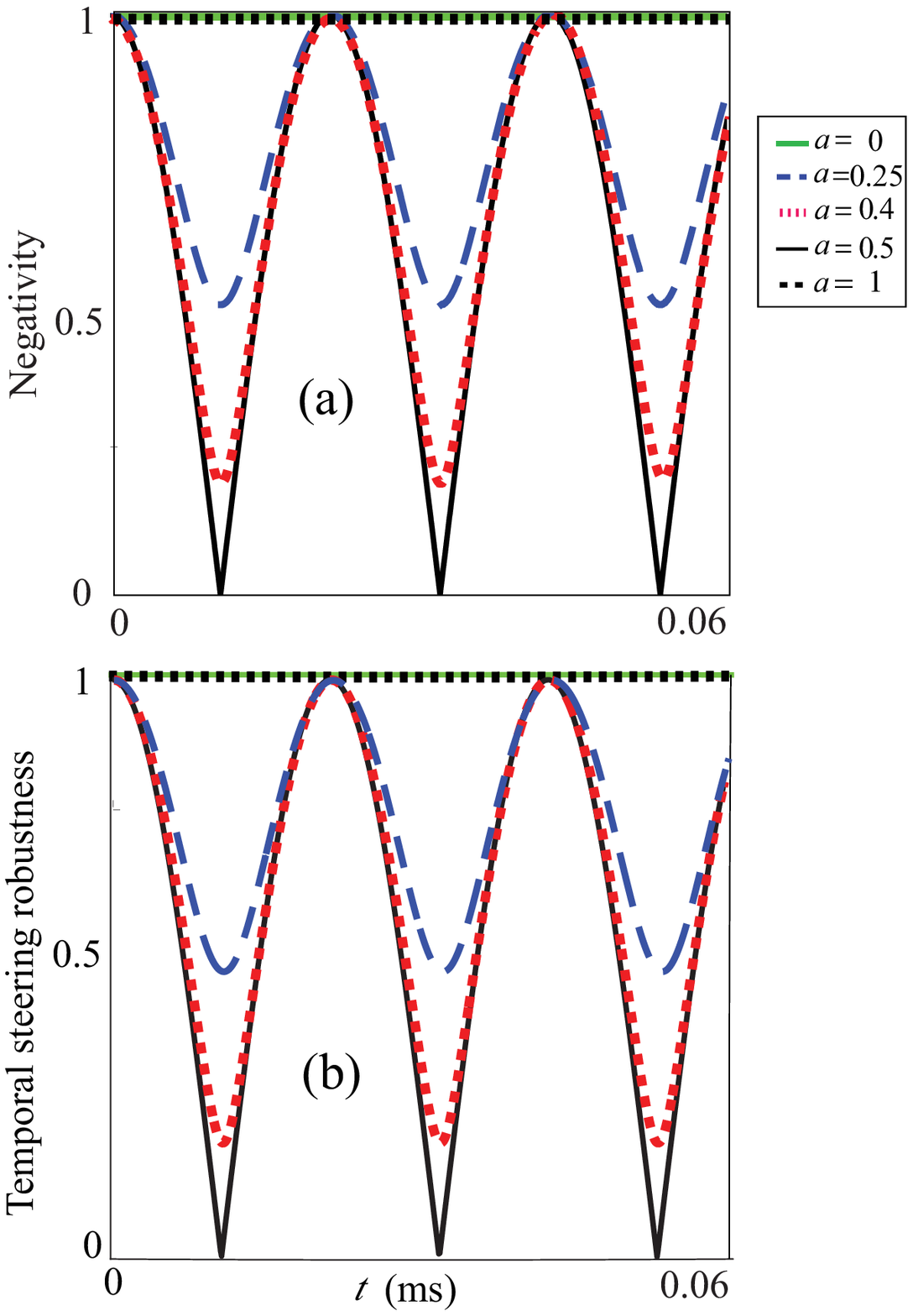}
\caption{(Color online)
The time evolution of the (a) negativity and (b) temporal steering robustness for two electrons of our simplified radical-pair model (i.e.,~$\Gamma=\kappa=0$).
The initial state is a direct product state between the electron singlet state and the nuclear spin state $\rho_{nu}(0)=a\mid\uparrow\rangle\langle\uparrow\mid+(1-a)\mid\downarrow\rangle\langle\downarrow\mid$, with the relative weight $a=0$~(green-solid), $a=0.25$~(blue-dashed), $a=0.4$~(red-dotted), $a=0.5$~(black-solid), and $a=1$ (black-dotted), respectively.
When $a=0.25$, $0.4$, and $0.5$, the oscillating curves indicate the non-Markovian nature of the dynamics.
In particular, for $a=0.5$, the nuclear spin possesses the largest Shannon entropy and results in the largest oscillation magnitudes in both panels.
Consequently, the dynamics of the two electrons shows the strongest non-Markovianity.
On the other hand, because $a=0$, $1$, the two electron state evolves unitarily. Hence, the negativity and temporal steering robustness are constant in time.
}
\label{negativity_different_a}
\end{figure}

In Ref.~\cite{Chen-S-L1}, it was shown that the temporal steerable weight is nonincreasing under completely positive and trace-preserving maps, hence it can be used to define a practical measure of non-Markovianity. Compare Eq.~\ref{TSR_SDP} with the SDP formulation of temporal steerable weight in Ref.~\cite{Chen-S-L1}; it is easy to show that temporal steering robustness can also reveal non-Markovian dynamics. The wavy curves in Figs.~\ref{TSR_mixed_setting_12} and \ref{TSR_mixed_setting_123} indicate the appearance of non-Markovianity in the radical-pair model. At first this may seem counterintuitive, because the equation of motion is in a Markovian Lindblad form, and,  when the hyperfine interaction tensor is $\mathbf{A}=\rm{diag}(0,0,\mathrm{A}_{\mathrm{z}})$, the nuclear-spin polarization remains unchanged during the spin dynamics~\cite{Cai2012}. However, because the initial state is assumed to be maximally mixed, the electrons effectively experience a mixture of two different evolutions, depending on the nuclear-spin state, leading to the observed non-Markovianity.

TO acquire more insights into this non-Markovianity, we simplify the model by neglecting the decay rate (i.e.,~$\Gamma=\kappa=0$) and consider the coherent dynamics of the two electrons and nuclear spin.
Assuming that the initial state is a direct product state between the electron singlet state and the nuclear-spin state $\rho_{nu}(0)=a|\uparrow\rangle\langle\uparrow|+(1-a)|\downarrow\rangle\langle\downarrow|$. The total density matrix at a later time can be expressed as
\begin{equation}
\rho(t)=a\rho^{1}_{e1,e2}(t)\otimes|\uparrow\rangle\langle\uparrow|+(1-a)\rho^{2}_{e1,e2}(t)\otimes|\downarrow\rangle\langle\downarrow|,
\label{dynamics_of_simplify_RDP}
\end{equation}
where 
\begin{equation}
\begin{aligned}
&\rho^{1}_{e1,e2}(t)=\exp[{i\mathrm{A}_{\mathrm{z}}\sigma_{\mathrm{z}}^{1}t}]|s\rangle\langle s|\exp[{-i\mathrm{A}_{\mathrm{z}}\sigma_{\mathrm{z}}^{1}t}]~~ \textrm{and}\\  
&\rho^{2}_{e1,e2}(t)=\exp[{-i\mathrm{A}_{\mathrm{z}}\sigma_{\mathrm{z}}^{1}t}]|s\rangle\langle s|\exp[{i\mathrm{A}_{\mathrm{z}}\sigma_{\mathrm{z}}^{1}t}],
\end{aligned}
\end{equation}
describe the dynamic evolutions of the two electrons under the influence of the magnetic fields locally induced by the nuclear spinors~\cite{Cai2012}.

\begin{figure}
\includegraphics[width=8cm]{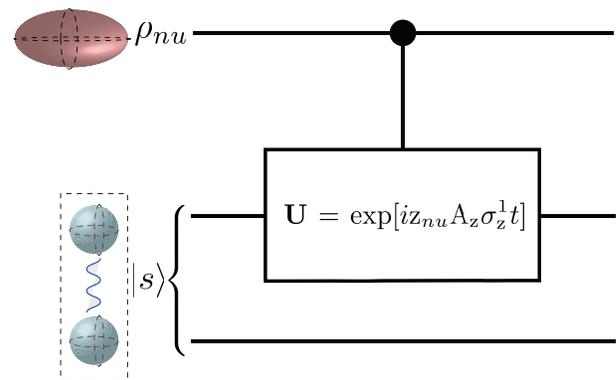}
\caption{(Color online) 
Schematic illustration revealing the analogy of our radical-pair model to a controlled-NOT gate.
The nuclear spin and two electrons play a role analogous to the control qubit and target qubit, respectively. The nuclear spin state decides the unitary operator $\mathbf{U}=\exp[{i\mathrm{z}_{nu} \mathrm{A}_{\mathrm{z}} \sigma_{\mathrm{z}}^{1}t}]$ exerting on electron 1, where $\mathrm{z}_{nu}=\pm 1$ is the eigenvalue of $\sigma_{\mathrm{z}}$. When $a$ gradually approaches $0.5$, the nuclear spin (C qubit) becomes more uncertain and possesses higher Shannon entropy. Therefore, the non-Markovianity of the two electrons is stronger.
}
\label{Fig_C-NOT_gate}
\end{figure}

To reveal the non-Markovian nature of the dynamics of the two electrons, we first notice that the state of the two electrons can be expressed as $\rho_{e1,e2}=\mathrm{Tr}_{nu}\rho(t)$.
Inspired by the RHP non-Markovianity measure~\cite{RHP2010}, in Fig.~\ref{negativity_different_a}, we show the entanglement of the two electrons quantified by the negativity~\cite{Horodecki_PRA1998}.
When $a=0$ or $1$, the nuclear spin is a pure state in $|\uparrow\rangle$ or $|\downarrow\rangle$, respectively, and the two electron state evolves unitarily. As the nuclear spin becomes a mixed state ($a=0.25$, 0.4, and 0.5), the two electron state is in the form of a convex combination of $\rho_{e1,e2}^{1}(t)$ and $\rho_{e1,e2}^{2}(t)$. Consequently, the time evolution of the entanglement between the two electrons shows oscillations.
Therefore, the nuclear spin plays the role of a non-Markovian environment. However, if we consider the entanglement of the nuclear spin and one of the electrons alone, by tracing out the other electron
, there is, of course, no entanglement between the nuclear spin and the electron~\cite{Cubitt2003}.

It is interesting to notice that the non-Markovianity of the convex combination of two unitary transformations, as given by Eq.~(\ref{dynamics_of_simplify_RDP}), can be seen as a pair of qubits coupled with each other via a controlled-NOT (CNOT) gate~\cite{Hong-Bin_PRA2015}. 
As shown in Fig.~\ref{Fig_C-NOT_gate}, the nuclear spin plays a role analogous to the control qubit (C qubit), which decides the corresponding mixture of unitary operators being exerted on electron 1.
It was shown in Ref.~\citep{Hong-Bin_PRA2015} that when $a$ gradually approaches $0.5$, namely the C qubit becomes more uncertain and possesses higher Shannon entropy, the target qubit exhibits stronger non-Markovianity.
This is exactly in line with our results that, as $a$ approaching $0.5$, the oscillation magnitude in Fig.~\ref{negativity_different_a} becomes larger, indicating stronger non-Markovianity in the dynamics of the two electrons.

\section{CONCLUSION}
In summary, we investigate the temporal steering robustness as a mean to quantify temporal steering in high-dimensional systems.
To explore its applications, we investigate the dynamics of temporal steering robustness in the radical-pair model. We show that the dynamics of the temporal steering robustness is clearly dependent on the orientation $\theta$.
We also reveal the non-Markovianity of the radical-pair model induced by the nuclear spin.
The time evolution of the radical pair is the convex combination of two unitary transformations. The different proportions of the nuclear state decide the convex combination of two unitary transformations of the radical pair.
When the nuclear spin state is up or down, the dynamics of the system is completely positive and trace-preserving.
However, when the nuclear spin is a mixed state, the radical pair behaves non-Markovianly.
It is interesting because the nuclear spins are in thermal equilibrium, a completely mixed state. It suggests that non-Markovianity not only plays a role in photosynthesis~\cite{Neill}, but may also have some influence in the avian compass.

\begin{acknowledgements}
This work is supported partially by the National Center for Theoretical Sciences and Ministry of Science and Technology, Taiwan, grant number MOST 103-2112-M-006-017-MY4.  We also acknowledge the support of a grant---from the John Templeton Foundation and also from a RIKEN-AIST collaboration grant. S.-L.C. acknowledges the support of
the DAAD/MOST Sandwich Program 2016 No. 57261473. Foundation. F.N. was partially supported by: the RIKEN iTHES Project,
the MURI Center for Dynamic Magneto-Optics via the AFOSR Award No. FA9550-14-1-0040, 
the Japan Society for the Promotion of Science (KAKENHI),
the ImPACT program of JST, and CREST.
\end{acknowledgements}
\section{APPENDIX}
In this appendix, we explicitly give the MUBs which are used as the measurement operators. The
MUBs are two orthonormal bases $\{|b_{1}\rangle,...|b_{d}\rangle\}$ and $\{|c_{1}\rangle,...|c_{d}\rangle\}$ of dimensions $d$, such that their complex inner-product between any basis states $|b_{i}\rangle$ and $|c_{j}\rangle$ can be expressed as $|\langle b_{i}|c_{j}\rangle|^{2}=1/d$~\cite{Klappenecker03}.
The set of $5$ MUBs is denoted by
$\{M_{a|x}\}_{a|x}$, with $a=1$,$2$,$3$,$4$; $x=1$,$2$,$3$,$4$,$5$; and $M_{a|x}=|\phi_{a|x}\rangle\langle\phi_{a|x}|$ where
\begin{equation} \label{MUBs}
\begin{split}
|\phi_{1|1}\rangle&=|1\rangle~~~~|\phi_{2|1}\rangle=|2\rangle\\
|\phi_{3|1}\rangle&=|3\rangle~~~~|\phi_{4|1}\rangle=|4\rangle\\
\\
|\phi_{1|2}\rangle&=\frac{1}{2}(|1\rangle+|2\rangle+|3\rangle+|4\rangle)\\
|\phi_{2|2}\rangle&=\frac{1}{2}(|1\rangle+|2\rangle-|3\rangle-|4\rangle)\\
|\phi_{3|2}\rangle&=\frac{1}{2}(|1\rangle-|2\rangle-|3\rangle+|4\rangle)\\
|\phi_{4|2}\rangle&=\frac{1}{2}(|1\rangle-|2\rangle+|3\rangle-|4\rangle)\\
\\
|\phi_{1|3}\rangle&=\frac{1}{2}(|1\rangle-|2\rangle-\mathrm{i}|3\rangle-\mathrm{i}|4\rangle)\\
|\phi_{2|3}\rangle&=\frac{1}{2}(|1\rangle-|2\rangle+\mathrm{i}|3\rangle+\mathrm{i}|4\rangle)\\
|\phi_{3|3}\rangle&=\frac{1}{2}(|1\rangle+|2\rangle+\mathrm{i}|3\rangle-\mathrm{i}|4\rangle)\\
|\phi_{4|3}\rangle&=\frac{1}{2}(|1\rangle+|2\rangle-\mathrm{i}|3\rangle+|4\rangle)\\
\\
|\phi_{1|4}\rangle&=\frac{1}{2}(|1\rangle-\mathrm{i}|2\rangle-\mathrm{i}|3\rangle-|4\rangle)\\
|\phi_{2|4}\rangle&=\frac{1}{2}(|1\rangle-\mathrm{i}|2\rangle+\mathrm{i}|3\rangle+|4\rangle)\\
|\phi_{3|4}\rangle&=\frac{1}{2}(|1\rangle+\mathrm{i}|2\rangle+\mathrm{i}|3\rangle-|4\rangle\\
|\phi_{4|4}\rangle&=\frac{1}{2}(|1\rangle+\mathrm{i}|2\rangle-\mathrm{i}|3\rangle+|4\rangle)\\
\\
|\phi_{1|5}\rangle&=\frac{1}{2}(|1\rangle-\mathrm{i}|2\rangle-|3\rangle-\mathrm{i}|4\rangle)\\
|\phi_{2|5}\rangle&=\frac{1}{2}(|1\rangle-\mathrm{i}|2\rangle+|3\rangle+\mathrm{i}|4\rangle)\\
|\phi_{3|5}\rangle&=\frac{1}{2}(|1\rangle+\mathrm{i}|2\rangle-|3\rangle+\mathrm{i}|4\rangle)\\
|\phi_{4|5}\rangle&=\frac{1}{2}(|1\rangle+\mathrm{i}|2\rangle+|3\rangle-\mathrm{i}|4\rangle).
\end{split}
\end{equation}
~~~~~~~~~~~~~~~~~~~~~~~~~~~~~~~~~~~~~~~~~~~



%

\end{document}